\documentclass[12pt]{article}
\usepackage{epsf,epsfig,float,amssymb,latexsym,amsmath,amsthm,fancyhdr,pst-all}
\usepackage{graphics,psfrag}
\newcommand{\xxxpc}{\%}
\newcommand{\la}{\langle}
\newcommand{\ra}{\rangle}

\newcommand{\be}{\begin{equation}}
\newcommand{\ee}{\end{equation}}
\newcommand{\ba}{\begin{eqnarray}}
\newcommand{\ea}{\end{eqnarray}}

\newcommand{\nn}{\nonumber}

\begin{document}

\thispagestyle{empty}

\vspace{2cm}

\begin{center}
{\Large{\bf Scalar radius of the pion and zeros in the form factor 
 }}
\end{center}
\vspace{.5cm}

\begin{center}
{\large Jos\'e A. Oller and Luis
Roca}
\end{center}

\begin{center}
{\it {\it Departamento de F\'{\i}sica. Universidad de Murcia.\\ E-30071,
Murcia. Spain.\\
{\small oller@um.es~,~luisroca@um.es}}}
\end{center}
\vspace{1cm}

\begin{abstract}
\noindent
The quadratic pion scalar radius, $\la r^2\ra^\pi_s$,  plays an important role for present precise 
determinations of $\pi\pi$ scattering. Recently,
 Yndur\'ain, using an Omn\`es representation of the null isospin(I)  non-strange pion scalar form 
 factor, obtains 
 $\la r^2\ra^\pi_s=0.75\pm 0.07$~fm$^2$. This value is  larger than the one calculated  
 by solving the corresponding 
  Muskhelishvili-Omn\`es equations, 
  $\la r^2\ra^\pi_s=0.61\pm 0.04$~fm$^2$. A large discrepancy between both values, 
  given the precision, then results. We reanalyze Yndur\'ain's method and show that by imposing 
  continuity of the resulting pion scalar form factor  under tiny changes in the input 
 $\pi\pi$ phase shifts, a zero in the form factor for  some S-wave I=0
  $T-$matrices is then required. 
   Once this is accounted for, the resulting value is $\la r^2\ra_s^\pi=0.65\pm 0.05$~fm$^2$.
     The main source of error in our determination is  
 present experimental uncertainties in low energy S-wave I=0 $\pi\pi$ phase shifts. Another 
 important contribution to our error is the not yet settled asymptotic behaviour of the 
 phase of the scalar form factor  from QCD. 
\end{abstract}

\vspace{2cm}


\newpage

\section{Introduction}
\label{sec:intro}
\def\theequation{\arabic{section}.\arabic{equation}}
\setcounter{equation}{0}

The scalar form factor of the pion, $\Gamma_\pi(t)$, corresponds to the matrix element
\be
\Gamma_\pi(t)=\int d^4 x \,e^{-i(q'-q)x}\la \pi(q')|\left(m_u \bar{u}(x)u(x)+ m_d \bar{d}(x)d(x)\right)
| \pi(q)\ra~,~~t=(q'-q)^2~.
\label{ffdef}
\ee

Performing a Taylor expansion around $t=0$,
\be
\Gamma_\pi(t)=\Gamma_\pi(0)\left\{1+\frac{1}{6}t\la r^2\ra_s^\pi+{\cal O}(t^2)\right\}~,
\label{r2pi}
\ee
where $\la r^2\ra_s^\pi$ is the quadratic scalar radius of the pion.

The quantity $\la r^2\ra_s^\pi$ contributes around 10$\xxxpc$ \cite{pipiscat} to the values of the 
S-wave $\pi\pi$ scattering lengths $a_0^0$ and $a_0^2$ as determined in ref.\cite{pipiscat}, 
by employing Roy equations and $\chi PT$ to two loops. If one takes into account that this reference
gives a precision of 2.2$\xxxpc$ in its calculation of the scattering lengths, a 10$\xxxpc$ 
of contribution from $\la r^2\ra_s^\pi$ is a large one.  Related to that, $\la r^2\ra_s^\pi$ 
is also important in $SU(2)\times SU(2)$ $\chi PT$ since it gives the 
low energy constant $\bar{\ell}_4$ that controls 
the departure of $F_\pi$ from its value in the chiral limit \cite{gl83,cd04} at leading order
correction.

Based on one loop $\chi PT$, Gasser and Leutwyler \cite{gl83} obtained $\la r^2\ra_s^\pi=0.55\pm
0.15$~fm$^2$. This calculation was improved later on by the same authors together with 
Donoghue \cite{dgl90}, who 
solved the corresponding Muskhelishvili-Omn\`es equations with the coupled 
channels of $\pi\pi$ and $K\bar{K}$.  The 
update of this calculation, performed in ref.\cite{pipiscat}, gives $\la r^2\ra_s^\pi=0.61\pm 0.04$
fm$^2$, where the new
results on S-wave I=0 $\pi\pi$ phase shifts from the Roy equation analysis of ref.\cite{acgl01} are
included.
 Moussallam \cite{m00} employs the same approach and obtains 
values in agreement with the previous result.

One should notice that solutions of the
 Muskhelishvili-Omn\`es equations for the scalar form
factor rely on non-measured $T-$matrix elements 
or on assumptions about which are the
channels that matter. 
Given the importance of $\la r^2\ra_s^\pi$, and the possible systematic errors 
in the analyses  based on Muskhelishvili-Omn\`es equations,
other independent approaches are most welcome.  In this respect we quote the works 
\cite{gu91,ou00,bct98}, and
 Yndur\'ain's ones \cite{y04,y05,y06}. 
These latter works have challenged the previous value for $\la r^2\ra_s^\pi$,
 shifting it to the larger $\la r^2\ra_s^\pi=0.75\pm 0.07$~fm$^2$. From ref.\cite{pipiscat} 
 the equations,
\be
\delta a_0^0 = +0.027 \Delta_{r^2}~,~
\delta a_0^2 = -0.004 \Delta_{r^2}~,
\ee
 give the change of the scattering lengths under a variation of $\la r^2\ra_s^\pi$
defined by 
$\la r^2\ra_s^\pi=0.61(1+\Delta_{r^2})$~fm$^2$. 
For the difference between the central values of $\la r^2\ra_s^\pi$ given above 
from refs.\cite{pipiscat,y04}, one has
$\Delta_{r^2}=+0.23$. This corresponds  to $\delta a_0^0=+0.006$ and $\delta a_0^2=-0.001$, 
while the errors  quoted are  $a_0^0=0.220\pm 0.005$ and $a_0^2=-0.0444\pm 0.0010$. We 
 then adduce about shifting the central values for the predicted scattering lengths at the level of one
sigma.

The value taken for $\la r^2\ra_s^\pi$ is also important for determining the ${\cal
O}(p^4)$ $\chi PT$ coupling 
$\bar{\ell}_4$. The value of ref.\cite{pipiscat} is $\bar{\ell}_4=4.4\pm 0.2$ while that  
of ref.\cite{y04} is $\bar{\ell}_4=5.4\pm 0.5$. Both values are incompatible within errors.

The papers \cite{y04,y05,y06} have been  questioned in refs.\cite{accgl05,ccl}. The
value of the $K\pi$ quadratic scalar radius, $\la r^2 \ra_s^{K\pi}$, obtained by Yndur\'ain 
in ref.\cite{y04}, $\la r^2 \ra_s^{K\pi}=0.31\pm0.06$~fm$^2$,  is not accurate,
 because he relies on old experiments and on a bad parameterization of low energy S-wave 
I=1/2 $K\pi$ phase shifts 
 by assuming dominance of the $\kappa$ resonance 
 as a standard Breit-Wigner pole \cite{opj04}.
  Furthermore, $\la r^2\ra_s^{K\pi}$ was recently fixed 
by high statistics experiments  in an interval in  agreement with the sharp prediction 
of \cite{opj04}, 
based on dispersion relations (three-channel Muskhelishvili-Omn\`es equations from the 
$T-$matrix of ref.\cite{opj00}) and two-loop 
$\chi$PT \cite{bt04}. From the  recent experiments \cite{istra,ktev}, one has 
 for the charged kaons \cite{istra} $\la r^2\ra_s^{K^\pm \pi}=0.235\pm 0.014\pm
 0.007$~fm$^2$, and for the neutral ones \cite{ktev} $\la r^2\ra_s^{K_L\pi}=0.165\pm 0.016$~fm$^2$. 
 The  prediction of \cite{opj04}, in an isospin limit,
  is $\la r^2\ra_s^{K\pi}=0.192\pm 0.012$~fm$^2$, 
 lying just in the middle of the experimental determinations. 
 Another issue is Yndur\'ain's  more sound determination of 
the pionic scalar radius, whose (in)correctness is not settled yet. 

In this paper we concentrate on the  approach of Yndur\'ain \cite{y04,y05,y06} to evaluate 
the quadratic scalar radius of the pion based on an Omn\'es representation of the  I=0 non-strange 
 pion scalar form factor. Our main conclusion will be that this approach 
\cite{y04} and the solution of the Muskhelishvili-Omn\`es equations \cite{dgl90}, with $\pi\pi$ and $K\bar{K}$ 
as coupled channels,  agree between each other if one properly takes into account, 
for some $T-$matrices, the
presence of a zero in the pion scalar form factor at energies slightly below the $K\bar{K}$ 
threshold. Precisely these $T-$matrices are 
those used in \cite{y04} and  favoured in \cite{y05}. Once this is considered 
 we conclude that
 $\la r^2\ra_s^\pi=0.63\pm 0.05$~fm$^2$.

The contents of the paper are organized as follows. In section 2 we discuss the Omn\`es representation of
$\Gamma_\pi(t)$ and derive the expression to calculate $\la r^2\ra_s^\pi$. This calculation is performed
in section 3, where we consider different parameterizations for experimental data and asymptotic phases for
the scalar form factor. Conclusions are given in the last section.

\section{Scalar form factor}
\label{sec:ff}
\def\theequation{\arabic{section}.\arabic{equation}}
\setcounter{equation}{0}

\begin{figure}[H]
\centerline{\epsfig{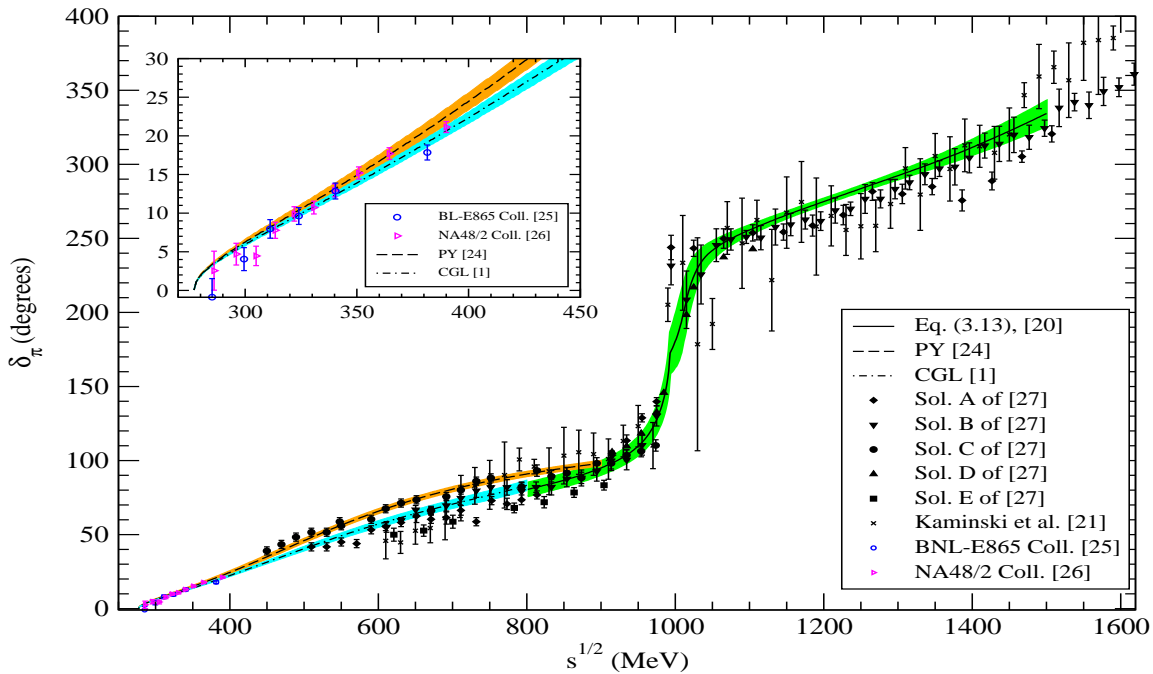}}
\vspace{0.2cm}
\caption[pilf]{\protect \small
S-wave $I=0$ $\pi\pi$ phase shift, $\delta_\pi(s)$.
 Experimental data are from refs.\cite{kaminski,bnl,na48,grayer}.
\label{figpi}}
\end{figure}

The pion scalar form factor $\Gamma_\pi(t)$, eq.(\ref{ffdef}),
 is an analytic function of $t$ with a
right hand cut, due to unitarity, for $t\geq 4 m_\pi^2$.
 Performing a dispersion relation of its 
logarithm,  with the possible zeroes of $\Gamma_\pi(t)$  removed, the Omn\`es 
representation results,
\be
\Gamma_\pi(t)=P(t)\exp\left[ \frac{t}{\pi} 
\int_{4m_\pi^2}^\infty \frac{\phi(s)}{s(s-t)}ds
 \right]~.
\label{ffomnes}
\ee
Here, $P(t)$ is a polynomial made up from the zeroes of $\Gamma_\pi(t)$,
with $P(0)=\Gamma_\pi(0)$. In the previous equation, $\phi(s)$ 
is the phase of $\Gamma_\pi(t)/P(t)$, taken to be continuous and such that 
$\phi(4m_\pi^2)=0$. In 
ref.\cite{y04} the scalar form factor is assumed to be free of zeroes and
hence $P(t)$ is just the constant $\Gamma_\pi(0)$ (the exponential 
factor is 1 for $t=0$). Thus,
\be
\Gamma_\pi(t)=\Gamma_\pi(0)\exp\left[ \frac{t}{\pi} \int_{4m_\pi^2}^\infty 
\frac{\phi(s)}{s(s-t)}ds \right]~.
\label{ffomnes2}
\ee
 From where it follows that,
\be
\la r^2\ra_s^\pi=\frac{6}{\pi}\int_{4 m_\pi^2}^\infty 
\frac{\phi(s)}{s^2}ds~.
\label{r2omnes1}
\ee 
One of the features of the pion scalar form factor of refs.\cite{dgl90,m00,ou00}, as 
discussed in ref.\cite{accgl05}, is the presence of a strong dip at energies around 
the $K\bar{K}$ threshold. This feature is also shared by the strong S-wave I=0 
$\pi\pi$ amplitude, $t_{\pi\pi}$. This is so because $t_{\pi\pi}$ is in very good 
approximation purely elastic below the $K\bar{K}$ threshold  and
 hence, neglecting inelasticity altogether in the discussion that follows,
 it is proportional to $\sin\delta_{\pi} e^{i\delta_\pi}$, with 
$\delta_\pi$ the S-wave I=0 $\pi\pi$ phase shift. It is an experimental fact that 
$\delta_\pi$ is very close to  $ \pi$ around the $K\bar{K}$ threshold, as shown in 
fig.\ref{figpi}. Therefore, if 
$\delta_\pi=\pi$ happens before the opening of this channel 
the strong amplitude has a zero at that energy. On the other hand, 
if $\delta_\pi=\pi$ occurs after the $K\bar{K}$ threshold, because inelasticity is then substantial, 
see eq.(\ref{tpipi}) below,
there is not a zero but a pronounced dip in $|t_{\pi\pi}|$.
 This dip can be arbitrarily close to zero if 
 before the $K\bar{K}$ 
threshold $\delta_\pi$ approaches $\pi$ more and more,  without reaching it. 

\begin{figure}[ht]
\centerline{\epsfig{file=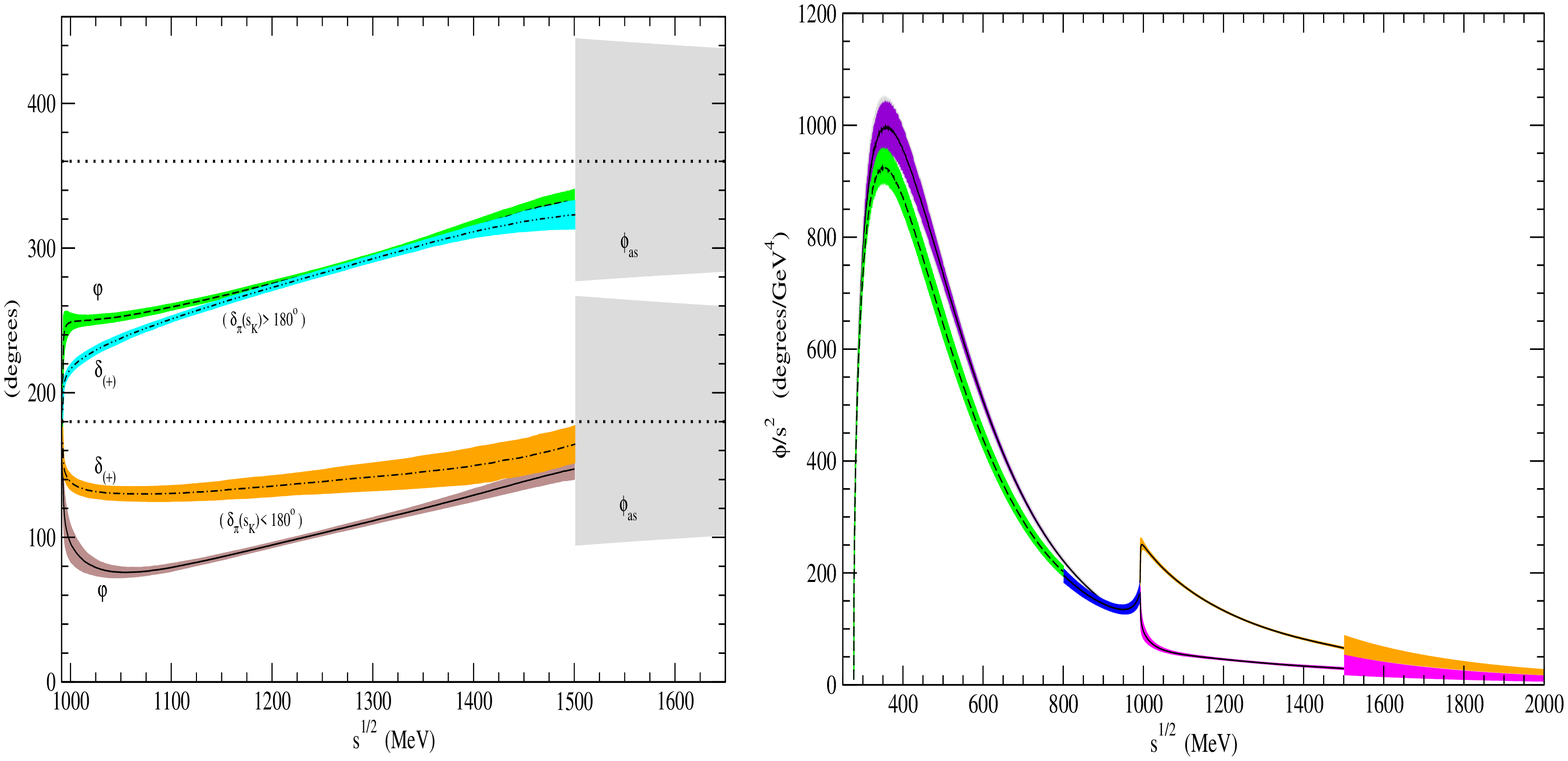,height=3.6in,width=7.in,angle=0}}
\vspace{0.2cm}
\caption[pilf]{\protect \small
Left panel: Strong phase $\varphi(s)$, eigenvalue phase $\delta_{(+)}(s)$ and asymptotic 
phase $\phi_{as}(s)$. Right panel: Integrand of $\la r^2\ra_s^\pi$ in  
eq.(\ref{split}) for parameterization I (dashed line) and II (solid line). For more details see the text. 
 Notice that the uncertainty due to $\phi_{as}(s)$ is much reduced in the
integrand.
\label{figpi2}}
\end{figure} 

\begin{figure}[ht]
\psfrag{ref}{\cite{paquito}}
\centerline{\epsfig{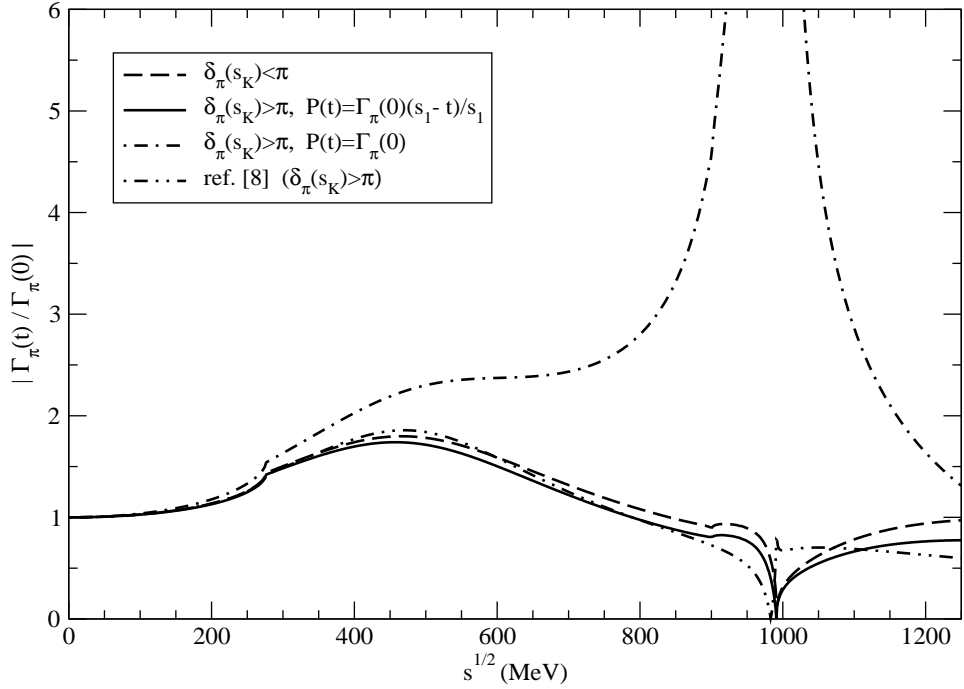}}
\vspace{0.2cm}
\caption[pilf]{\protect \small
$|\Gamma_\pi(t)/\Gamma_\pi(0)|$ from eq.(\ref{ffomnes2})
 with $\delta_\pi(s_K)<\pi$, dashed-line, 
 and $\delta_\pi(s_K)>\pi$, dashed-dotted line. The solid line corresponds to use 
    eq.(\ref{ffomnes3}) for the latter case. For this figure 
  we have used parameterization II (defined in section \ref{sec:resul}) with
 $\alpha_1=2.28$ (dashed line) and 2.20 (dashed-dotted and solid lines). 
  The dashed-double-dotted line is  the scalar form factor
of ref.\cite{ou00} that has $\delta_\pi(s_K)>\pi$.
\label{figpi3}}
\end{figure}

Because of Watson final state theorem the phase $\phi(s)$ in eq.(\ref{ffomnes}) 
is given by $\delta_\pi(s)$ below 
the $K\bar{K}$ threshold, neglecting inelasticity due to $4\pi$ or $6\pi$ states  
 as indicated by experiments \cite{hyams}.  The situation above the $K\bar{K}$ threshold 
is more involved. Let us recall that 
\be
t_{\pi\pi}=(\eta \,e^{2i\delta_\pi}-1)/2i~,
\label{tpipi}
\ee 
with $0\leq \eta \leq 1$ and   
the inelasticity is given by $1-\eta^2$, with $\eta$ the elasticity coefficient. We denote 
by $\varphi(s)$ the phase of $t_{\pi\pi}$, required to be continuous (below $4m_K^2$ it is 
given by $\delta_\pi(s)$). By continuity, 
 close enough to the $K\bar{K}$ threshold and above it, $\eta\to 1$ and then 
we are in the same situation as in the elastic case. As a result, because of the Watson final state
theorem and continuity, the phase $\phi(s)$ must still be given by $\varphi(s)$. 
   For $\delta_\pi(s_K)<\pi$, $s_K=4m_K^2$,   $\varphi(s)$ does not follow the increasing 
trend with energy of $\delta_\pi(s)$ but drops as a result of 
eq.(\ref{tpipi}), see fig.\ref{figpi2} for $\delta_\pi(s_K)<\pi$. 
This is easily seen  by writing explicitely the real and imaginary 
parts of $t_{\pi\pi}$ in eq.(\ref{tpipi}),
\be
t_{\pi\pi}=\frac{1}{2}\eta\sin2\delta_\pi+\frac{i}{2}(1-\eta\cos 2\delta_\pi)~.
\label{tpipi2}
\ee
The imaginary part is always positive ($\eta<1$ above the $K\bar{K}$ threshold and 1.1 GeV \cite{hyams})
 while the real part is negative for $\delta_\pi<\pi$,  but in an interval of just a few MeV  
  the real part turns positive as 
 soon as $\delta_\pi>\pi$, fig.\ref{figpi}. As a result, 
 $\varphi(s)$ passes quickly from values below but close to $\pi$ to the interval 
 $[0,\pi/2]$. This rapid motion of $\phi(s)$ 
 gives rise to a pronounced minimum of $|\Gamma_\pi(t)|$ at this energy, as 
indicated in ref.\cite{accgl05} and shown in fig.\ref{figpi3}.
 The drop in $\phi(s)$ becomes more and more dramatic as $\delta_\pi(s_K)\to \pi^-$ (with the superscript 
$+(-)$ indicating that the limit is approached from values above(below), respectively); 
 and in this limit, $\phi(s_k)=\varphi(s_K)$ is
 discontinuous at $s_K$. This is easily understood from eq.(\ref{tpipi2}). Let us call $s_1$ the point
 at which $\delta_\pi(s_1)=\pi$ with $s_1>s_K$.  Close and above $s_1$,
  $\varphi(s)\in [0,\pi/2]$, for
 the reasons explained above, and 
$\varphi(s)$ has decreased very rapidly from almost $\pi$ at the $K\bar{K}$ threshold to 
values below $\pi/2$ just after $s_1$. Then, in the limit  $s_1\to s_K^+$ one has 
  $\phi(s_K^-)=\varphi(s_K^-)=\pi$ on the left, while on the right 
  $\phi(s_K^+)=\varphi(s_K^+)<\pi/2$.
  As a result $\varphi(s)$ is discontinuous at $s=s_K$.  
We stress that this discontinuity of $\varphi(s)$ at $s_K$
 when $\delta_\pi(s_K)\to\pi^-$ applies rigorously to $\phi(s_K)$ as well  
since $\eta(s_K)=1$. This discontinuity  at 
$s=s_K$ implies also that the integrand in the Omn\`es representation for $\Gamma_\pi(t)$ 
develops a logarithmic singularity as,
\be
\frac{\phi(s_K^-)-\phi(s_K^+)}{\pi}\log\frac{\delta}{s_K}~,
\ee
with $\delta\to 0^+$.
When exponentiating this result  one has 
a zero for $\Gamma_\pi(s_K)$ as $(\delta/s_K)^\nu$, $\nu=(\phi(s_K^-)-\phi(s_K^+))/\pi>0$ and 
$\delta\to 0^+$. 
This zero is a necessary consequence when evolving continuously from $\delta_\pi(s_K)<\pi$ to 
$\delta_\pi(s_K)>\pi$.\footnote{It can be shown from eq.(\ref{tpipi2}) that 
$\phi(s_K^-)-\phi(s_K^+)=\pi$. Here we are assuming $\eta=1$ for $s\leq s_K$, which is a very good
 approximation as indicated by experiment \cite{hyams,kaminski}.} This in turn implies rigorously that 
in the Omn\`es representation of $\Gamma_\pi(t)$,
 eq.(\ref{ffomnes}), $P(t)$ must be a polynomial of first degree for those 
 cases with $\delta_\pi(s_K)\geq \pi$,\footnote{We are focusing in the physically relevant region of 
 experimental
 allowed values for $\delta_\pi(s_K)$, which  can be larger or smaller than $\pi$ but close to.}
 \be
 P(t)=\Gamma_\pi(0)\frac{s_1-t}{s_1}~,
 \ee
 with $s_1$ the position of the zero. Notice that the degree of the polynomial $P(t)$ is 
 discrete and thus by continuity it cannot change unless a singularity develops. This is the 
case when $\delta_\pi(s_K)=\pi$, changing the degree from 0 to 1. 
  Hence, if $\delta_\pi(s_K)\geq \pi$ for a given $t_{\pi\pi}$, instead of eqs.(\ref{ffomnes2}) and (\ref{r2omnes1}) one must then consider,
 \be
\Gamma_\pi(t)=\Gamma_\pi(0)\frac{s_1-t}{s_1}\exp\left[\frac{t}{\pi}\int_{4m_\pi^2}^\infty 
\frac{\phi(s)}{s(s-t)}ds\right]~,
\label{ffomnes3}
 \ee
 and
\be
\la r^2\ra_s^\pi=-\frac{6}{s_1}+\frac{6}{\pi}\int_{4m_\pi^2}^\infty
\frac{\phi(s)}{s^2}ds~.
\label{r2omnes2}
\ee 
For those $t_{\pi\pi}$ for which $\delta_\pi(s_K)>\pi$ then 
 $\varphi(s)$ follows $\delta_\pi(s)$ just after the $K\bar{K}$ threshold and there is no drop, 
as emphasized in ref.\cite{y05}, see fig.\ref{figpi2}. 

Summarizing,   we have shown that $\Gamma_\pi(t)$ has a zero at $s_1$ when
 $\delta_\pi(s_K)\geq \pi$ as a consequence of the assumption  that $\phi(s)$ follows $\varphi(s)$ 
 above the $K\bar{K}$ threshold, along the lines of ref.\cite{y05}, and by imposing continuity 
 in $\Gamma_\pi(t)$ under small changes in  $\delta_\pi(s_K)\simeq \pi$.  As a result 
 eqs.(\ref{ffomnes3}) and (\ref{r2omnes2}) should be used in the latter case, instead of 
 eqs.(\ref{ffomnes2}) and (\ref{r2omnes1}), valid for $\delta_\pi(s_K)<\pi$. 
  This solution was overlooked in refs.\cite{y04,y05,y06}. We show in appendix~A 
 why the previous discussion  on the zero of $\Gamma_\pi(t)$ 
 for $\delta_\pi(s_K)\geq \pi$ at $s_1$ 
  cannot be applied to all pion scalar form factors, in particular to the strange one.

If eq.(\ref{ffomnes2}) were used for those $t_{\pi\pi}$ with $\delta_\pi(s_K)
\geq \pi$ then a strong maximum of $|\Gamma_\pi(t)|$ would be  obtained 
around the $K\bar{K}$ threshold, 
instead of the aforementioned zero or the minimum of refs.\cite{dgl90,m00}, as shown
 in fig.\ref{figpi3} by the dashed-dotted line.
That is also shown in fig.10 of ref.\cite{paquito} or fig.2 of \cite{accgl05}. This is the situation
 for the $\Gamma_\pi(t)$ of refs.\cite{y04,y05}, and it is the reason why 
 $\la r^2\ra_s^\pi$ obtained there is much larger than that of refs.\cite{dgl90,pipiscat,m00}. 
 That is, Yndur\'ain uses
 eqs.(\ref{ffomnes2}), (\ref{r2omnes1}) for 
 $\delta_\pi(s_K)\geq \pi$,  instead of  
 eqs.(\ref{ffomnes3}), (\ref{r2omnes2})  (solid line in fig.\ref{figpi3}). 
 The unique and important role 
played by $\delta_\pi(s_K)$ (for elastic $t_{\pi\pi}$ below the $K\bar{K}$ threshold)
 is perfectly recognised in ref.\cite{y05}. However, in this reference
the astonishing conclusion that $\Gamma_\pi(t)$ has two radically different behaviours 
under tiny variations of $t_{\pi\pi}$ was sustained. These  variations are enough to pass 
from $\delta_\pi(s_K)<\pi$ to $\delta_\pi(s_K)\geq \pi$ \cite{y04}, while the $T-$ or 
$S-$matrix are fully continuous. Because of this instability of the solution of refs.\cite{y04,y05} 
  under tiny 
changes of $\delta_\pi(s)$, we consider ours, that produces continuous
 $\Gamma_\pi(t)$,
  to be certainly preferred. We also stress that our solutions, either for $ \delta_\pi(s_K)\geq \pi$ and 
$\delta_\pi(s_K)<\pi$, are the ones that agree with those obtained by solving  the  
 Muskhelishvili-Omn\`es equations \cite{dgl90,pipiscat,m00} and Unitary $\chi$PT \cite{ou00}.

Let us now show how to fix $s_1$ in terms of the knowledge  of 
  $\delta_\pi(s)$ with $\delta_\pi(s_K)\geq \pi$.
   For this purpose let us perform a dispersion relation 
   of $\Gamma_\pi(t)$ with two subtractions,
\be
\Gamma_\pi(t)=\Gamma_\pi(0)+\frac{1}{6}\la r^2\ra_s^\pi t+\frac{t^2}{\pi}\int_{4m_\pi^2}^\infty
\frac{\hbox{Im}\Gamma_\pi(s)}{s^2(s-t)}ds~,
\label{ffdis1}
\ee
 From asymptotic QCD \cite{brodsky} one expects that the scalar form factor vanishes at infinity \cite{y04,y06}, then the dispersion integral in 
eq.(\ref{ffdis1}) should converge rather fast. Eq.(\ref{ffdis1}) is useful 
because it tells us that the only point around 1 GeV where there can be a
 zero in $\Gamma_\pi(t)$ is at the energy $s_1$ for which the imaginary part of $\Gamma_\pi(t)$ vanishes. 
 Otherwise, the integral in the right hand side of eq.(\ref{ffdis1}) 
 picks up an imaginary part and there is 
 no way to cancel it  as $\Gamma_\pi(0)$, $\la r^2\ra_s^\pi$ and $t$ are all real.
  Since $|\hbox{Im}\Gamma_\pi(t)|=|\Gamma_\pi(t)\,\sin\delta_\pi(t)|$ for $t\leq s_K$,
it  certainly vanishes at the point 
 $s_1$ where $\delta_\pi(s_1)=\pi$. As there is only one zero at such energies, 
 this determines $s_1$ exactly in terms of 
 the given parameterization for $\delta_\pi(s)$. 

One could argue against the  argument just given to determine $s_1$ that this energy could be 
complex. However, this would imply two zeroes at $s_1$ and $s_1^*$, and then the degree of $P(t)$ would 
be two instead of one. Notice that the degree of the polynomial $P(t)$ is 
 discrete and thus, by softness in the continuous parameters of the $T-$matrix, 
 its value should stay at 1 for some open domain in the parameters with 
$\delta_\pi(s_K)>\pi$ until a discontinuity develops. Physically, the presence of two zeroes 
would in
turn require that $\phi(s)\to 3\pi$ 
 so as to guarantee that $\Gamma_\pi(t)$ still vanishes as $-1/t$, as 
 required by asymptotic QCD \cite{brodsky,y04}. This value for the asymptotic
  phase seems to be rather unrealistic as $\varphi(s)$ only reaches $2\pi$ 
  at already quite high energy values, as shown in fig.\ref{figpi2}.

\section{Results}
\label{sec:resul}

Our main result from the previous section is the sum rule to determine $\la r^2\ra_s^\pi$,
\be
\la r^2\ra_s^\pi=-\frac{6}{s_1}\theta(\delta_\pi(s_K)-\pi)+
\frac{6}{\pi}\int_{4m_\pi^2}^\infty\frac{\phi(s)}{s^2}ds~,
\label{r2final}
\ee
where  $\theta(x)=0$ for $x<0$ and 1 for $x\geq 0$. We 
 split $\la r^2\ra_s^\pi$ in two
parts:
\ba
\la r^2\ra_s^\pi&=&Q_H+Q_A~,\nn\\
Q_H&=&-\frac{6}{s_1}\theta(\delta_\pi(s_K)-\pi)+\frac{6}{\pi}\int_{4m_\pi^2}^{s_H} \frac{\phi(s)}{s^2}ds~,\nn\\
Q_A&=&\frac{6}{\pi}\int_{s_H}^\infty \frac{\phi(s)}{s^2}~,
\label{split}
\ea
with $s_H=2.25$ GeV$^2$. Reasons for fixing $s_H$ to this value are given below.

The main issue in the application of 
eq.(\ref{r2final})  is to determine $\phi(s)$ in the 
 integrand.  Below the $K\bar{K}$ threshold and neglecting inelasticity, 
 one has that $\phi(s)=\delta_\pi(s)$, $4m_\pi^2\leq s \leq 4 m_K^2$. 
This follows because of 
 the Watson final state theorem, continuity and the equality 
 $\phi(4m_\pi^2)=\delta_\pi(4m_\pi^2)=0$. 
 
 For practical applications we shall  
consider the S-wave I=0 $\pi\pi$ phase shifts given by
 the $K-$matrix parameterization of ref.\cite{hyams} (from its energy dependent
analysis of  data from 0.6 GeV up to 1.9 GeV) and the
 parameterizations of ref.\cite{pipiscat} (CGL) and  ref.\cite{py03} (PY). The resulting
$\delta_\pi(s)$ for all these parameterizations are shown in fig.\ref{figpi}. We use 
 CGL from $\pi\pi$ threshold up to 0.8 GeV, because this is the upper limit of 
its analysis, while PY is used up to 0.9 GeV, because at this energy 
it matches  well inside the experimental errors with the data of \cite{hyams}. 
The $K-$matrix of ref.\cite{hyams} is used for energies above 0.8 GeV, when using 
 CGL below this energy (parameterization I), and above 0.9 GeV, when using PY 
 for lower energies (parameterization II). We take 
the parameterizations  CGL and PY  as their difference below 0.8 GeV accounts 
well for the experimental uncertainties in $\delta_\pi$, see fig.\ref{figpi}, and 
 they satisfy constraints from $\chi PT$ (the former) and dispersion relations (both). The reason why we
skip to use the parameterization of ref.\cite{hyams} for lower energies is because one should be 
there as
precise as possible since this region gives the largest contribution to $\la r^2\ra_s^\pi$,  
 as it is evident from the right panel of fig.\ref{figpi2}. 
 It happens that the $K-$matrix of \cite{hyams}, that fits data above 0.6 GeV,
 is not compatible with data from $K_{e 4}$ decays 
\cite{bnl,na48}. We show in the insert of fig.\ref{figpi} the comparison of the parameterizations 
 CGL and PY with the $K_{e4}$ data of \cite{bnl,na48}. 
 We also show in the same figure the experimental points on $\delta_\pi$ 
from refs.\cite{hyams,kaminski,grayer}. 
Both refs.\cite{hyams,kaminski} are compatible within errors, with some disagreement 
above 1.5 GeV. 
 This disagreement does not affect our numerical results since above 1.5 GeV we do not rely on data.

The $K-$matrix of ref.\cite{hyams} is given by,
\be
K_{ij}(s)=\alpha_i\alpha_j/(x_1-s)+\beta_i\beta_j/(x_2-s)+\gamma_{ij}~,
\label{km}
\ee
 where
 \be
 \begin{array}{lll}
 x_1^{1/2}=0.11\pm 0.15~ & x_2^{1/2}=1.19\pm 0.01 & \\
\alpha_1=2.28\pm 0.08~ & \alpha_2=2.02\pm 0.11 & \\
\beta_1=-1.00\pm 0.03~ & \beta_2=0.47\pm 0.05 &\\
\gamma_{11}=2.86\pm 0.15~ & \gamma_{12}=1.85\pm 0.18~ & \gamma_{22}=1.00\pm 0.53~,
 \end{array}
 \label{array}
 \ee
with units given in appropriate powers of GeV. 
In order to calculate the contribution from the phase shifts of this $K-$matrix
 we generate Monte-Carlo gaussian samples, taking into account the errors shown in eq.(\ref{array}),
and evaluate $Q_H$ according to eq.(\ref{split}).
The central value of $\delta_\pi(s_K)$ for the 
 $K-$matrix of ref.\cite{hyams} is $3.05$, slightly below $\pi$. When 
generating Monte-Carlo gaussian samples  according to eq.(\ref{array}), 
there are cases with $\delta_{\pi}(s_K)\geq \pi$, around $30\xxxpc$ of the samples. 
 Note that for these cases one also has 
the contribution $-6/s_1$ in eq.(\ref{r2final}).

The application of  Watson final state theorem for $s>4m_K^2$ is not straightforward 
 since inelastic channels are relevant.
The first important one is the $K\bar{K}$ channel associated in turn with 
the appearance of the narrow $f_0(980)$ resonance, just on
top of its threshold. This implies a sudden drop of the elasticity parameter 
$\eta$, but it again rapidly raises (the $f_0(980)$ resonance is narrow with a width around 30 MeV) 
and in  the region $1.1^2\lesssim s\lesssim 1.5^2$ GeV$^2$
 is compatible within errors 
with $\eta=1$ \cite{hyams,kaminski}. For $\eta\simeq 1$, 
 the Watson final state theorem would imply
again that $\phi(s)=\varphi(s)$, but, as emphasized by \cite{accgl05}, this equality 
 only holds, in principle, modulo $\pi$. The reason advocated in ref.\cite{accgl05} is 
 the presence of the region $s_K<s<1.1^2$ GeV$^2$ where  inelasticity can be 
large, and then 
continuity arguments alone cannot be applied 
to guarantee the equality $\phi(s)\simeq \varphi(s)$ for $s\gtrsim 1.1^2$~GeV$^2$. 
This argument has been proved in ref.\cite{y05} to be quite irrelevant in the present case.
In order to show this a diagonalization of the $\pi\pi$ and $K\bar{K}$ $S-$matrix is done. 
These channels  
 are the relevant ones when $\eta$ is clearly different from 1, between 1 and 1.1 GeV.
 Above that energy one also has the opening of the $\eta\eta$ channel and the increasing role of 
 multipion states.
  
  We reproduce here the arguments of ref.\cite{y05}, but deliver expressions directly 
 in terms of the phase shifts and elasticity parameter, instead of $K-$matrix parameters 
  as done in ref.\cite{y05}. For two channel scattering, because of unitarity, 
 the $T-$matrix can be written as:
\be
T=\left(
\begin{array}{ll}
\frac{1}{2i}(\eta e^{2i\delta_\pi}-1) & \frac{1}{2}\sqrt{1-\eta^2}e^{i(\delta_\pi+\delta_K)} \\
\frac{1}{2}\sqrt{1-\eta^2}e^{i(\delta_\pi+\delta_K)} & \frac{1}{2i}(\eta e^{2i \delta_K}-1)
\end{array}
\right)~,
\label{tmat}
\ee
with $\delta_K$ the elastic S-wave I=0 $K\bar{K}$ phase shift. In terms of
the $T$-matrix the S-wave I=0 $S-$matrix is given by,
\be
S=I+2i T~,
\label{smat}
\ee
satisfying $S S^\dagger=S^\dagger S =I$. The $T$-matrix can also be written 
as 
\be
T=Q^{1/2}\left(K^{-1}-i  Q\right)^{-1} Q^{1/2}~,
\label{tmat2}
\ee
where the $K-$matrix is real and symmetric  along the real axis for $s\geq 4m_\pi^2$ and 
$Q=diag(q_\pi,q_K)$, with $q_\pi(q_K)$ the center of mass momentum of pions(kaons).
 This allows
one to diagonalize $K$ with a real orthogonal matrix $C$, and hence both the
$T-$ and $S-$matrices are also diagonalized with the same matrix. Writing,
\be
C=\left(
\begin{array}{cc}
\cos\theta & \sin \theta \\
-\sin \theta & \cos \theta
\end{array}
\right)~,
\label{cmat}
\ee 
one has
\ba
\cos\theta &=&\frac{\left[(1-\eta^2)/2\right]^{1/2}}{\left[
1-\eta^2\cos^2\Delta-\eta|\sin\Delta|\sqrt{1-\eta^2\cos^2\Delta}\right]^{1/2}}~,\nn \\
\sin\theta &=&-\frac{\sin\Delta}{\sqrt{2}}\frac{\eta-\sqrt{1+(1-\eta^2)\cot^2\Delta}}{\left[
1-\eta^2\cos^2\Delta-\eta|\sin\Delta|\sqrt{1-\eta^2\cos^2\Delta}\right]^{1/2}}~,
\label{costeta}
\ea
with $\Delta=\delta_K-\delta_\pi$. On the other hand, 
 the eigenvalues of the $S-$matrix are given by,
\ba
e^{2i \delta_{(+)}}&=&S_{11}\frac{1+e^{2i\Delta}}{2}
\left[1-\frac{i}{\eta}\tan\Delta\, 
\sqrt{1+(1-\eta^2)\cot^2\Delta}\right]\\
e^{2i \delta_{(-)}}&=&S_{22}\frac{1+e^{-2i\Delta}}{2}
\left[1+\frac{i}{\eta}\tan\Delta\,
 \sqrt{1+(1-\eta^2)\cot^2\Delta}\right]~.
 \label{eigenvalues}
\ea
The eigenvalue phase $\delta_{(+)}$ satisfies $\delta_{(+)}(s_K)=\delta_\pi(s_K)$. 
 The expressions above for $\exp 2i\delta_{(+)}$
 and $\exp 2i\delta_{(-)}$ interchange between each other when $\tan\Delta$ crosses zero and 
 simultaneously the sign in the right hand side of eq.(\ref{costeta}) for $\sin\theta$ changes.  
This diagonalization allows to disentangle two elastic scattering channels. The scalar form
factors attached to every of these channels,  $\Gamma'_1$ and $\Gamma'_2$, 
will satisfy the Watson final state theorem in the whole
energy range and then one has,
\ba
\Gamma'&\equiv& \left(
\begin{array}{c}
\Gamma'_1\\
\Gamma'_2
\end{array}\right)
=C^T Q^{1/2}\Gamma=C^T Q^{1/2}\left(
\begin{array}{c}
\Gamma_\pi \\
\Gamma_K
\end{array}\right)~,\nn\\
\Gamma_\pi &=& q^{-1/2}_\pi\left(
\lambda \cos \theta \,|\Gamma_1'| e^{i\delta_{(+)}} \pm  \sin \theta \, |\Gamma_2'|
e^{i\delta_{(-)}}\right)~,\nn\\
\Gamma_K &=& q_K^{-1/2}\left(
 \pm \cos \theta \, |\Gamma_2'|
e^{i\delta_{(-)}} - \lambda \sin \theta\, |\Gamma_1'| e^{i\delta_{(+)}}\right)~.
\label{diag}
\ea
The $\pm$ in front of $|\Gamma'_2|$ is due to the fact that $\Gamma'_2=0$ at $s_K$, as 
follows from its definition in the equation above. Since Watson final state theorem only fixes 
the phase of $\Gamma'_2$ up to  modulo $\pi$, and the phase is not defined in the zero, 
 we  cannot fix the sign in front  at this stage. Next,  $\Gamma'_1$ has a zero at $s_1$  
 when $\delta_\pi(s_K)\geq \pi$. For this case,   
$-|\Gamma_1'|$ must appear in the previous equation, so as to guarantee continuity
of its ascribed phase, and this is why 
$\lambda=(-1)^{\theta(\delta_\pi(s_K)-\pi)}$.

Now, when $\eta\to 1$ then $\sin\theta\to 0$ as $\sqrt{(1-\eta)/2}$ and 
$\phi(s)$ is then the eigenvalue phase $\delta_{(+)}$. This eigenvalue phase
can be calculated given the $T-$matrix. For those $T-$matrices 
employed here, and those of
refs.\cite{y04,y05,dgl90,accgl05}, $\delta_{(+)}(s)$
 follows rather closely  $\varphi(s)$ in the whole energy range. This is shown in  fig.\ref{figpi2} 
 and already discussed in detail in ref.\cite{y05}. In this way, one guarantees 
 that $\phi(s)$ and $\varphi(s)$  do not differ between each other  
in an integer multiple of $\pi$ when $\eta\simeq 1$, $1.1^2\lesssim s \lesssim 1.5^2$~GeV$^2$.

For the calculation of $Q_H$ in eq.(\ref{split}) we shall equate $\phi(s)=\varphi(s)$ 
for $4m_K^2<s<1.5^2$ GeV$^2$. Denoting,
\ba
I_H&=&\frac{6}{\pi}\int_{4m_\pi^2}^{s_H}\frac{\varphi(s)}{s^2}= I_1+I_2+I_3~,\nn\\
I_1&=&\frac{6}{\pi}\int_{4m_\pi^2}^{s_K} \frac{\varphi(s)}{s^2}ds~,\nn\\
I_2&=&\frac{6}{\pi}\int_{s_K}^{1.1^2} \frac{\varphi(s)}{s^2}ds~,\nn\\
I_3&=&\frac{6}{\pi}\int_{1.1^2}^{s_H} \frac{\varphi(s)}{s^2}ds~,
\label{ies}
\ea
then 
\be 
Q_H\simeq I_H-\frac{6}{s_1}\theta(\delta_\pi(s_K)-\pi)~.
\label{qhfinal}
\ee
Now, eq.(\ref{diag}) can also be used to estimate the error of approximating 
$\phi(s)$ by $\varphi(s)$ in the range $4m_K^2< s < 1.5^2$ GeV$^2$  to calculate 
$I_2$ and $I_3$ as done in eq.(\ref{ies}). 
We could have also used $\delta_{(+)}(s)$ in eq.(\ref{ies}). However, 
notice that when $\eta\lesssim 1$ then $\varphi(s)\simeq \delta_{(+)}(s)$ 
and when inelasticity could be substantial the difference between $\delta_{(+)}(s)$ and $\varphi(s)$  
 is well taken into account in the error analysis that follows. Remarkably, consistency 
 of our approach also requires
 $\phi(s)$ to be closer to $\varphi(s)$ than to $\delta_{(+)}(s)$. The reason is that 
 $\varphi(s)$ for $\delta_\pi(s_K)\geq \pi$ is in very good approximation the $\varphi(s)$ for 
$\delta_\pi(s_K)<\pi$ plus $\pi$, this is clear from fig.\ref{figpi2}. This  difference 
is $precisely$ the required one in order to have the same value for $\la r^2\ra_s^\pi$ either for 
$\delta_\pi(s_K)<\pi$ or $\delta_\pi(s_K)\geq \pi$ from eq.(\ref{r2final}). However, the difference for 
 $\delta_{(+)}(s)$ between $\delta_\pi(s_K)<\pi$ and $\delta_\pi(s_K)\geq \pi$ is smaller than 
 $\pi$. Indeed, we note that $\phi(s)$ follows closer $\varphi(s)$ than $\delta_{(+)}(s)$ 
 for the explicit form factors of refs.\cite{ou00,dgl90}. 
 
Let us consider first the range $1.1^2<s<1.5^2$ GeV$^2$ where from
experiment \cite{hyams} $\eta\simeq 1$ within
errors.  With $\epsilon=\pm \tan\theta |\Gamma_2'/\Gamma_1'|$ and
$\rho=\delta_{(-)}-\delta_{(+)}$, eq.(\ref{diag}) allows us to write,
\ba
\Gamma_\pi=\lambda \cos\theta\,|\Gamma_1'|e^{i\delta_{(+)}}(1+\epsilon \cos\rho)
\left(1+i\frac{\epsilon\sin\rho}{1+\epsilon \cos\rho}\right)~.
\label{phase1}
\ea
When $\eta\to 1$  then $\epsilon\to 0$, according to the  expansion,\footnote{The  the ratio 
$\left|\Gamma_2'/\Gamma_1'\right|$, present in $\epsilon$, is not expected to 
be large since the $f_0(1300)$ couples mostly to $4\pi$ and similarly to $\pi\pi$ and 
$K\bar{K}$, and the $f_0(1500)$ does mostly to $\pi\pi$ \cite{pdg}.}
\be
\tan\theta=\frac{\sqrt{(1-\eta)/2}}{\sin\Delta}\left[1
-\frac{1+3\cos 2\Delta}{8\sin^2\Delta}(1-\eta)\right]
+{\cal O}\left((1-\eta)^{5/2}\right)~.
\ee
Rewriting, 
\be
1+i\frac{\epsilon\sin\rho}{1+\epsilon \cos\rho}=\exp\left(
i\frac{\epsilon\sin\rho}{1+\epsilon \cos\rho}\right)+{\cal O}(\epsilon^2)~,
\label{phase2}
\ee
which  from eqs.(\ref{phase1}) and (\ref{phase2}) implies a shift in $\delta_{(+)}$ because 
of inelasticity effects,
\be
\delta_{(+)}\to \delta_{(+)}+\frac{\epsilon\sin\rho}{1+\epsilon \cos\rho}~.
\label{epsi}
\ee

Using $\eta=0.8$ in the range $1.1^2\lesssim s\lesssim 1.5^2$ GeV,  $\eta\simeq 1$ 
from the energy dependent analysis of
ref.\cite{hyams} given by the $K-$matrix of eq.(\ref{km}),
one ends with $\epsilon\simeq 0.3$. Taking into account that $\delta_{(+)}$ 
is larger than $\gtrsim 3\pi/2$ for  $\delta_\pi(s_K)\geq \pi$ (in this case 
$\delta_{(+)}\simeq \delta_\pi$), and around $3\pi/4$ for 
$\delta_\pi(s_K)<\pi$, see fig.\ref{figpi2}, one ends with 
relative corrections to $\delta_{(+)}$ around $6\xxxpc$ for the
former case and $13\xxxpc$ for the latter. 
Although the $K-$matrix of ref.\cite{hyams}, eq.(\ref{km}), is given up to 1.9 GeV, one should be aware
 that to take only the two channels $\pi\pi$ and $K\bar{K}$
 in the whole energy range is an oversimplification, 
particularly above 1.2 GeV.  Because of this we  finally double the 
previous
estimate. Hence $I_3$ is calculated with a relative error 
of $12\xxxpc$ for $\delta_\pi(s_K)\geq \pi$ and $25\xxxpc$ for $\delta_\pi(s_K)<\pi$.

In the narrow region between $s_K<s<1.1^2$ GeV$^2$,  $\eta$ can be rather different from 1, 
due to the $f_0(980)$ that couples
very strongly to the just open $K\bar{K}$ channel. 
However, from the direct measurements 
of $\pi\pi\to K\bar{K}$ \cite{expipikk}, where $1-\eta^2$ is directly measured,\footnote{Neglecting
multipion states.} 
 one has a better way to determine $\eta$ than from $\pi\pi$ scattering
\cite{hyams,kaminski}. It results from the former  
experiments, as shown also by explicit calculations \cite{npa,nd,ao07},  that $\eta$ is not so small as indicated
in $\pi\pi$ experiments \cite{hyams},  and one has
$\eta\simeq 0.6-0.7$ for its minimum value. Employing $\eta=0.6$ in eq.(\ref{epsi}) then   
$\epsilon\simeq 0.5$. Taking $\delta_{(+)}$ around $\pi/2$ when
$\delta_\pi(s_K)<\pi$ this implies a relative error of 30$\xxxpc$. For 
$\delta_\pi(s_K)\geq \pi$ one has instead $\delta_{(+)}\gtrsim \pi$, and a 
$15\xxxpc$ of estimated error. Regarding the ratio of the
moduli of form factors entering in $\epsilon$ we expect it to be $\lesssim 1$ (see appendix A). 
 Therefore, our error in the evaluation of 
$I_2$ is estimated to be $30\xxxpc$ and $15\xxxpc$ for the cases 
$\delta_\pi(s_K)<\pi$ and $\delta_\pi(s_K)\geq \pi$, respectively.

As a result of the discussion following eq.(\ref{qhfinal}), we
 consider that the error estimates done for $I_2$ and $I_3$ in the case $\delta_\pi(s_K)<\pi$ are too
 conservative and that the  relative errors 
  given for $\delta_\pi(s_K)>\pi$ are more realistic. Nonetheless, since 
the absolute errors that one obtains for $I_2$ and $I_3$ are 
 the same in both cases (because $I_2$ and $I_3$ for $\delta_\pi(s_K)<\pi$ are around a factor 
 2 smaller than those for $\delta_\pi(s_K)\geq \pi$) we keep the errors as given above. 
 To the previous errors for $I_2$ and $I_3$ due to inelasticity, we also add in quadrature the noise 
in the calculation of $Q_H$ due to the error in $t_{\pi\pi}$ from the uncertainties in 
the parameters of the $K-$matrix eqs.(\ref{km}), (\ref{array}), and those in the 
parameterizations CGL and PY. 

We finally employ for $s>2.25$ GeV$^2$ the knowledge of the asymptotic phase 
  of the pion scalar form factor in order  to evaluate $Q_A$ in eq.(\ref{split}). 
 The function $\phi(s)$ is determined so as to match with the asymptotic 
 behaviour of $\Gamma_\pi(t)$ 
as $-1/t$ from QCD. 
The  Omn\`es representation of the scalar form factor, 
eqs.(\ref{ffomnes2}) and (\ref{ffomnes3}), tends to $t^{-q/\pi}$ and $t^{-q/\pi+1}$ for 
$t\to\infty$, respectively. Here, $q$ is the asymptotic value of the phase 
$\phi(s)$ when $s\to \infty$. Hence, for $\delta_\pi(s_K)<\pi$ the function $\phi(s)$ 
 is then required 
to tend to $\pi$ while for $\delta_\pi(s_K)\geq \pi$ the asymptotic value should be $2\pi$. 
The way $\phi(s)$ is predicted to approach the limiting value is somewhat ambiguous 
\cite{y05,y06},

\be
\phi_{as}(s)\simeq \pi\left( n\pm\frac{2d_m}{\log(s/\Lambda^2)}\right)~.
\label{asin}
\ee
In this equation, $2d_m=24/(33-2 n_f)\simeq 1$, $\Lambda^2$
is the QCD scale parameter
 and $n=1,~2$ for $\delta_\pi(4m_K^2)<\pi,~\geq \pi$, respectively. 
 The case $n=2$ was not discussed  in refs.\cite{y04,y05,y06,accgl05,ccl} 
 for the form factor given in eq.(\ref{ffdef}).
 There is as well a controversy between \cite{ccl} and \cite{y06} regarding the $\pm$ sign
  in eq.(\ref{asin}). If leading twist contributions dominate 
\cite{y05,y06} then the limiting value is reached from above and one has the plus sign, while if 
twist three contributions are the dominant ones \cite{ccl} the minus sign  has to be considered \cite{y06}.
 In the left panel of fig.\ref{figpi2} we show with the wide bands 
the values of $\phi(s)_{as}$ for $s>2.25$ GeV$^2$ from eq.(\ref{asin}), considering both signs,
 for $n=1$ ($\delta_\pi(s_K)<\pi$) and $2$ ($\delta_\pi(s_K)\geq \pi$). 
  We see in the figure that above $1.4-1.5$ GeV ($1.96-2.25$ GeV$^2$)
  both $\varphi(s)$ and $\phi(s)_{as}$ phases match and this is why we take 
  $s_H=2.25$ GeV$^2$ in eq.(\ref{r2final}), similarly as done in refs.\cite{y04,y05}. 
 In this way, we also avoid to enter into hadronic details in a region where $\eta<1$ with 
  the onset of the $f_0(1500)$ resonance.   
 The present uncertainty whether the $+$ or $-$ sign holds in  eq.(\ref{asin}) is 
taken as a source of error in evaluating $Q_A$. The other source of uncertainty comes
from the value taken for $\Lambda^2$, $0.1 <\Lambda^2<0.35$ GeV$^2$, as suggested in ref.\cite{y04}.
 From fig.\ref{figpi2} it is clear that our error 
 estimate for $\phi_{as}(s)$ is very conservative and should account
  for uncertainties due to the onset of inelasticity for energies 
  above $1.4-1.5$ GeV and to the appearance of the $f_0(1500)$ resonance. In the right 
panel of fig.\ref{figpi2} we show the integrand for $\la r^2\ra_s^\pi$, 
eq.(\ref{split}), for parameterization I (dashed line) and II (solid line).
 Notice as the large uncertainty in $\phi_{as}(s)$ is much reduced in the
integrand as it happens for the higher energy domain.

\begin{table}
\begin{center}
\begin{tabular}{|c|c|c|c|c|}
\hline
$\phi(s)$ & I & I & II  & II \\
\hline
$\delta_\pi(s_K)$ & $\geq \pi$ & $<\pi$ &  $\geq \pi$ & $<\pi$ \\ \hline
$I_1$ &  $0.435\pm 0.013$   & $0.435\pm 0.013$ & $0.483\pm 0.013 $& $0.483\pm 0.013 $ \\
$I_2$ &  $0.063\pm 0.010$   & $0.020\pm 0.006$ & $0.063\pm 0.010 $& $0.020\pm 0.006 $ \\
$I_3$ &  $0.143\pm 0.017$   & $0.053\pm 0.013 $ & $0.143\pm 0.017$ & $0.053\pm 0.013 $ \\
$Q_H$ &  $0.403\pm 0.024$ & $0.508\pm 0.019$ & $0.452\pm 0.024 $& $0.554\pm 0.019 $ \\
$Q_A$ &  $0.21 \pm 0.03$   & $0.10\pm 0.03$ & $0.21\pm 0.03 $& $0.10\pm 0.03 $ \\
\hline 
$\la r^2\ra_s^\pi$ &  $0.61\pm 0.04$   & $0.61\pm 0.04$ & $0.66\pm 0.04$& $0.66\pm 0.04 $ \\
\hline
\end{tabular}
\caption{Different contributions to $\la r^2\ra_s^\pi$ as defined in
 eqs.(\ref{split}) and (\ref{ies}). All the units are $\textrm{fm}^2$.
In the value for $\la r^2\ra_s^\pi$ the errors due to $I_1$, $I_2$, $I_3$ and $Q_A$ are added in quadrature.
\label{tableresul}}
\end{center}
\end{table}

In table \ref{tableresul} we show the values of $I_1$, $I_2$, $I_3$, $Q_H$, $Q_A$ and $\la r^2\ra_s^\pi$
  for the parameterizations 
I and II and for the two cases $\delta_\pi(s_K)\geq \pi$ and $\delta_\pi(s_K)<\pi$. 
 This table shows the disappearance of the disagreement 
between the cases $\delta_\pi(s_K)\geq \pi$ and $\delta_\pi(s_K)<\pi$ from the $\pi\pi$ and $K\bar{K}$ 
$T-$matrix of eq.(\ref{km}), once the zero of $\Gamma_\pi(t)$ at $s_1<s_K$ is 
taken into account for the former case.  This disagreement was the reason for the controversy
 between Yndur\'ain 
and ref.\cite{accgl05} regarding the value of $\la r^2\ra_s^\pi$. 
The fact that the parameterization II gives rise to
 a larger value of $\la r^2\ra_s^\pi$ than I is because PY follows the upper $\delta_\pi$ data 
 below 0.9 GeV, while CGL follows  lower ones, as shown in fig.\ref{figpi}.

The different errors in table \ref{tableresul} are added in quadrature. 
The final value for $\la r^2\ra_s^\pi$ is the mean between those of parameterizations 
I and II and the error is taken such that it spans 
the interval of values in table \ref{tableresul} at the level of
two sigmas. One ends with:
\ba
\la r^2\ra_s^\pi=0.63\pm 0.05~\hbox{fm}^2~.
\label{values}
\ea
The largest sources of error in $\la r^2\ra_s^\pi$ are the uncertainties in the experimental 
$\delta_\pi$ and in the asymptotic phase $\phi_{as}$. This is due to the fact that the former are enhanced because 
of its weight in the integrand, see fig.\ref{figpi2}, and the latter due to its large size. 

Our number above and  that of refs.\cite{pipiscat,dgl90}, $\la r^2\ra_s^\pi=0.61\pm 0.04$~fm$^2$, 
are then compatible. 
On the other hand, we have also evaluated $\la r^2\ra_s^\pi$ directly from 
the scalar form factor obtained with
 the dynamical approach of ref.\cite{ou00} from Unitary $\chi$PT and we obtain
$\la r^2\ra_s^\pi=0.64\pm 0.06$ fm$^2$, in perfect agreement with eq.(\ref{values}). 
Notice that the scalar 
form factor of ref.\cite{ou00} has $\delta_\pi(s_K)>\pi$ and we have checked that it has a 
zero at $s_1$, as it should. This is shown in fig.\ref{figpi3} by the dashed-double-dotted line. The value  
$\la r^2\ra_s^\pi=0.75\pm 0.07$~fm$^2$ from refs.\cite{y04,y05} is much larger than ours because 
the possibility of a zero at $s_1$ was not taking into account there and 
other solution was considered. This solution, however, 
has an unstable behaviour under the transition $\delta_\pi(s_K)=\pi-0^+$ to 
$\delta_\pi(s_K)=\pi+0^+$ and it 
cannot be connected continuously with the one for  $\delta_\pi(s_K)<\pi$. 
Our solution for $\Gamma_\pi(t)$ from Yndur\'ain's method does not have this unstable behaviour
 and it is continuous under changes in the 
values of the parameters of the $K-$matrix, eqs.(\ref{km}) and (\ref{array}). This is why, 
from our results, it follows too that the interesting discussion of ref.\cite{y05}, regarding 
whether $\delta_\pi(s_K)<\pi$ or $\geq \pi$, is not any longer conclusive 
to explain the disagreement between the values of refs.\cite{y04,y05} and 
 ref.\cite{pipiscat} for $\la r^2\ra_s^\pi$.

We can also work out from our determination of $\la r^2\ra_s^\pi$, eq.(\ref{values}), values for the 
${\cal O}(p^4)$ $SU(2)$ $\chi PT$ low energy constant $\bar{\ell}_4$. We take the two loop 
expression in $\chi PT$ for $\la r^2 \ra_s^\pi$ \cite{pipiscat}, 
\be
\la r^2 \ra_s^\pi=\frac{3}{8\pi^2 f_\pi^2}\left\{ \bar{\ell}_4-\frac{13}{12}+\xi \Delta_r\right\}~,
\label{oneloop}
\ee
where $f_\pi=92.4$ MeV is the pion decay constant,
 $\xi=(M_\pi/4\pi f_\pi)^2$ and $M_\pi$ is the pion mass.
 First, at the one loop level calculation $\Delta_r=0$ and then one obtains, 
\ba
\bar{\ell}_4&=&4.7\pm 0.3 ~.
\ea 
We now move to the determination of $\bar{\ell}_4$ based on the full two loop 
relation between $\la r^2\ra_s^\pi$ and $\bar{\ell}_4$.
The expression for $\Delta_r$ can be found in Appendix~C of ref.\cite{pipiscat}. $\Delta_r$
 is given
in terms of one ${\cal O}(p^6)$ $\chi PT$ counterterm, $\widetilde{r}_{S_2}$, 
and four ${\cal O}(p^4)$ ones. Taking the values of all these parameters, but for $\bar{\ell}_4$, from 
ref.\cite{pipiscat}, and solving for $\bar{\ell}_4$, one arrives to 
\ba
\bar{\ell}_4&=&4.5\pm 0.3~.
\label{l4values}
\ea
 This  number is in good agreement with $\bar{\ell}_4=4.4\pm 0.2$ \cite{pipiscat}.
 
 Ref.\cite{y06} also points out that  one loop $\chi$PT fits to the S-, P- and D-wave scattering lengths and effective
 ranges give rise to much larger values for $\bar{\ell}_2$ and $\bar{\ell}_4$ than those of
 ref.\cite{pipiscat}. For more details we refer to \cite{y06}. 
\section{Conclusions}
\label{sec:conclu}

In this paper we have addressed the issue of the discrepancies between the values 
of the quadratic pion scalar radius of Leutwyler {\it et al.} \cite{dgl90,accgl05}, 
$\la r^2\ra_s^\pi=0.61\pm 0.04$~fm$^2$, and   
 Yndur\'ain's papers \cite{y04,y05,y06}, $\la r^2\ra=0.75\pm 0.07$~fm$^2$. 
 One of the reasons of interest for having a precise determination of $\la r^2\ra_s^\pi$ 
is its contribution of a 10$\xxxpc$ to $a_0^0$ and $a_0^2$, calculated with a precision 
of $2\xxxpc$ in ref.\cite{pipiscat}. The value taken for $\la r^2\ra_s^\pi$ is also important for
determining the ${\cal O}(p^4)$ $\chi$PT coupling $\bar{\ell}_4$. 

 From our study it follows that Yndur\'ain's method to calculate $\la r^2\ra_s^\pi$
 \cite{y04,y05}, based on an Omn\`es representation of the pion scalar form factor,
 and that derived by solving the two(three) coupled channel Muskhelishvili-Omn\`es equations 
 \cite{dgl90,pipiscat,m00},
are compatible. It is shown that the reason for the aforementioned discrepancy is the presence of 
a zero in $\Gamma_\pi(t)$ for those S-wave I=0 $T-$matrices with $\delta_\pi(s_K)\geq \pi$ and elastic below 
the $K\bar{K}$ threshold, with $s_K=4 m_K^2$. 
This zero was overlooked in refs.\cite{y04,y05},
 though, if one imposes continuity in the 
 solution obtained under tiny changes of the $\pi\pi$ phase shifts employed, 
 it is necessarily required by the approach followed there. 
 Once this zero is taken into account the same value for $\la r^2\ra_s^\pi$ is obtained irrespectively 
 of whether $\delta_\pi(s_K)\geq \pi$ or $\delta_\pi(s_K)<\pi$. Our final result is $\la r^2\ra_s^\pi=
 0.63\pm 0.05$~fm$^2$.
  The error estimated takes into account experimental uncertainty in the values of 
 $\delta_\pi(s)$, inelasticity effects and present ignorance in the way the phase of the form 
 factor approaches its asymptotic value $\pi$, as predicted from
 QCD.   
 Employing our value for $\la r^2\ra_s^\pi$ we calculate 
 $\bar{\ell}_4=4.5\pm 0.3$. The values $\la r^2\ra_s^\pi=0.61\pm 0.04$~fm$^2$ and 
 $\bar{\ell}_4=4.5\pm 0.3$ of ref.\cite{pipiscat} are then in good agreement with ours.

\section*{Acknowledgements}
 We thank Miguel Albaladejo for providing us  numerical results from some 
unpublished $T-$matrices and Carlos Schat for his collaboration in a parallel research.
 We also thank F.J. Yndur\'ain for long discussions and B. Anathanarayan, I. Caprini, 
 G. Colangelo, J. Gasser and H. Leutwyler for a critical reading of a previous 
 version of the manuscript. This work was supported in part by the MEC (Spain) and FEDER (EC) Grants
  FPA2004-03470 and Fis2006-03438,  the 
  Fundaci\'on  S\'eneca (Murcia) grant Ref. 02975/PI/05, the European Commission
(EC) RTN Network EURIDICE under Contract No. HPRN-CT2002-00311 and the HadronPhysics I3
Project (EC)  Contract No RII3-CT-2004-506078.

\section*{Appendices}

\appendix

\section{Coupled channel dynamics}
\def\theequation{A.\arabic{equation}}
\setcounter{equation}{0}
\label{appen:1}

We take $\pi\pi$ and $K\bar{K}$ coupled channels and denote by $F_1$ and $F_2$ their respective 
I=0 scalar form factors. Unitarity requires,
\be
 \hbox{Im}F_i=\sum_{j=1}^2 F_j \rho_j \theta(t-s'_j) t_{ji}^*~,
\ee 
where $||t_{ij}||$ is the I=0 S-wave $T-$matrix, $s'_i$ is the threshold energy square of channel $i$ and 
$\rho_i=q_i/8\pi\sqrt{s}$, with $q_i$ its center of mass
three momentum.

A general solution to the previous equations is given by,
\be
F=T \,G~,~F=\left(\begin{array}{c} F_1 \\ F_2 \end{array}\right)~,~
G=\left(\begin{array}{c}G_1 \\ G_2\end{array}\right)~,
\ee
where the functions $G_i(t)$ do not have right hand cut. This equation is interesting as tells us that if
pion dynamics dominate, $|G_1|>>|G_2|$, then $F_1\simeq G_1 t_{11}$ and the form factor phase $\phi(s)$ 
follows $\varphi(s)$. As a result, like $t_{11}$, it has a zero at $s_1$ below the $K\bar{K}$ threshold for 
$\delta_\pi(s_K)\geq \pi$, as shown in section \ref{sec:resul}. 
On the other hand, if kaon dynamics dominates, $|G_2|>>|G_1|$, then $F_1\simeq G_2t_{12}$ and 
 $\phi(s)$ follows the phase of $t_{12}$, that above the $K\bar{K}$ 
threshold is clearly above $\pi$. This is why for the pion strange scalar form factor 
there is no zero at $s_1\lesssim s_K$ for $\delta_\pi(s_K)\geq \pi$, indeed there is a 
maximum like that shown in fig.\ref{figpi3} by the dashed-dotted line.

 As in section \ref{sec:resul} 
we now proceed to the diagnolization above the $K\bar{K}$ threshold of the
 renormalized $T-$matrix $T'$,
 \ba
&& T'=\rho^{1/2} T \rho^{1/2}~,~
 \rho=\left(\begin{array}{c}\rho_1^{1/2} \\ \rho_2^{1/2}\end{array}\right)~,~
\widetilde{T}=C^T T' C=\left(\begin{array}{cc}\widetilde{t}_{11} & 0 \\ 
0 & \widetilde{t}_{22}\end{array}\right)~,\nn\\
&&\widetilde{t}_{11}=\sin \delta_{(+)}e^{i\delta_{(+)}}~,~
\widetilde{t}_{22}=\sin \delta_{(-)}e^{i\delta_{(-)}}~.
\ea
The corresponding diagonal form factors $F'_1$ and $F'_2$, collected in the vector $F'$, are
\ba
F'&=&C^T \rho^{1/2} F=\widetilde{T}C^T\rho^{-1/2}G=\left(\begin{array}{ll}
\left\{\cos\theta \,\rho_1^{-1/2} G_1-\sin\theta \, \rho^{-1/2}_2 G_2\right\}\widetilde{t}_{11}\\
\left\{\sin\theta\, \rho_1^{-1/2}G_1+\cos \theta\, \rho_2^{-1/2}G_2\right\}\widetilde{t}_{22}
\end{array}\right)~.
\label{fprima}\ea
The previous expressions allow to obtaining  $F_1$ directly in terms of the
eigenphases and with clean separation between pion, $G_1$, and kaon dynamics, $G_2$. From
eq.(\ref{diag}) it follows that,
\ba
\begin{aligned}
F_1=&\left\{\cos^2\theta\,\rho^{-1}G_1-\cos\theta\,\sin \theta\,\rho_2^{-1/2}\rho_1^{-1/2}G_2\right\}
\widetilde{t}_{11}\\
+&\left\{\sin^2\theta \,\rho_1^{-1}G_1+\cos\theta\,\sin\theta\,\rho_2^{-1/2}\rho_1^{-1/2}G_2\right\}
\widetilde{t}_{22}~.
\end{aligned}
\label{f1f1}
\ea
For $\delta_\pi(s_K)\geq \pi$ typical values, somewhat above the $K\bar{K}$ threshold, are
 $e^{2 i\delta_{(+)}}\simeq +i$, $e^{2i\delta_{(-)}}\simeq -i$ and 
 $\sin\theta>0$. For dominance of $G_1$ one has 
 $F_1/G_1\simeq \rho_1^{-1}(i+\cos 2\theta)/2$ while for dominance of $G_2$ the result is 
 $F_1/G_2\simeq -\sin\theta\,\cos\theta\,\rho_2^{-1/2}\rho_1^{-1/2}<0$. The factors $G_{1,2}$ do
  not introduce any change
 in $\phi(s)$ with respect to its value before the opening of the $K\bar{K}$ threshold 
 since they are smooth functions in $s$.\footnote{Due to the Adler zeroes this is not necessarily case 
 close to the $\pi\pi$ threshold.}
  In both cases the phase $\phi(s)$ is larger than $\pi$ 
 and $F_1$ follows the upper trend of phases shown in fig.\ref{figpi2} 
 (note that in this case $\widetilde{t}_{11}$
  is in the first quadrant though $\delta_\pi>\pi$). Now, doing the same exercise for
 $\delta_\pi(s_K)<\pi$, one has the typical values $e^{2 i\delta_{(+)}}\simeq -i$,
  $e^{2i\delta_{(-)}}\simeq +i$ and  $\sin\theta<0$. For pion dominance then 
$F_1/G_1\simeq \rho_1^{-1}(i-\cos2\theta)/2$ and for the kaon one
 $F_1/G_2\simeq +\sin\theta \cos\theta \rho_2^{-1/2} 
\rho_1^{-1/2}<0$. Thus, in the former case the phase is $\gtrsim \pi/2$, and follows the
lower trend of phases of fig.\ref{figpi2}, while in the 
latter is $\gtrsim \pi$ and follows again the upper trend (this is the case of the strange scalar form
factor).  

The demonstration in section \ref{sec:resul} that $\phi(s_K)$ is discontinuous in the limit
$\delta_\pi(s_K)\to\pi^-$ by taking $s_1\to s_K^+$, 
cannot be applied in the case of kaon dominance (e.g. pion strange scalar form factor). From eq.(\ref{f1f1}) 
it follows that, 
\be
F_1(t)\simeq -\cos\theta\sin\theta \rho_2^{-1/2}\rho_1^{-1/2}
G_2\left(\widetilde{t}_{11}-\widetilde{t}_{22}\right)~.
\ee
The point is that  $\widetilde{t}_{22}$ for $t\geq s_1$ ($s_1\to s_K^+$) is of size  
comparable with that of $\widetilde{t}_{11}$ (both tend to zero) and the phase does not follow  
 $\delta_{(+)}$. This is
not the case for pion dominance because for $s_1\to s_K^+$ 
then $\sin^2\theta\to 0$, $F_1(t)\simeq
\cos^2\theta \,\rho_1^{-1} G_1 \widetilde{t}_{11}$, eq.(\ref{f1f1}), and $\phi(s)$ follows $\delta_{(+)}$.

 From eq.(\ref{fprima}) we can also write  $|\Gamma'_2/\Gamma'_2|\simeq |\widetilde{t}_{11}
\tan\theta/\widetilde{t}_{22}|$ for the case of pion dominance. Since typically
$|\widetilde{t}_{11}/\widetilde{t}_{22}|\simeq 1$,
 as shown above for energies somewhat above the $K\bar{K}$ threshold, 
then  $|\Gamma'_2/\Gamma'_1|\simeq |\tan\theta|<1$. This is why we consider that 
equating it to 1 in section \ref{sec:resul} is a conservative estimate.

\end{document}